# Experimental interference of uncorrelated photons

Heonoh Kim[1], Osung Kwon[2,†] & Han Seb Moon[1,*]

[1]Department of Physics, Pusan National University, Geumjeong-Gu, Busan 46241, South Korea

[2]Affiliated Institute of Electronics and Telecommunications Research Institute, Daejeon 34044, South Korea

†E-mail: oskwon@nsr.re.kr

*Corresponding author: hsmoon@pusan.ac.kr

**Abstract**

The distinguishing of the multiphoton quantum interference effect from the classical one forms one of the most important issues in modern quantum mechanics and experimental quantum optics. For a long time, the two-photon interference (TPI) of correlated photons has been recognized as a pure quantum effect that cannot be simulated with classical lights. In the meantime, experiments have been carried out to investigate the classical analogues of the TPI. In this study, we conduct TPI experiments with uncorrelated photons with different center frequencies from a luminescent light source, and we compare our results with the previous ones of correlated photons. The observed TPI fringe can be expressed in the form of three phase terms related to the individual single-photon and two-photon states, and the fringe pattern is strongly affected by the two single-photon-interference fringes and also by their visibilities. With the exception of essential differences such as valid and accidental coincidence events within a given resolving time and the two-photon spectral bandwidth, the interference phenomenon itself exhibits the same features for both correlated and uncorrelated photons in the single-photon counting regime.

**Introduction**

From the early 17th century, the interference of light has formed one of the most important topics in natural science[1]. In modern quantum mechanics, the interference of photons is at the heart of understanding the superposition of quantum states and correlations between distinct particles[2,3]. From Young's two-slit experiment on the classical interference phenomenon[4] to modern quantum-optical experiments utilizing correlated photons[5-8], a consistent understanding of multiphoton interference effects has played a key role in realizing new photonic quantum information technologies[9-14].

Over the past 30 years, the Hong-Ou-Mandel (HOM) interference effect[15] has been considered as the most representative multiphoton quantum interference phenomenon because it cannot be treated using the classical wave theory; further, there is no classical counterpart that entirely mimics the fringe visibility and the related two-photon state. Moreover, the observation of the HOM effect is essential to understand the superposition principle in fundamental quantum mechanics and the implementation of quantum information processing[3,10,13]. Since the first report of the HOM experiment, a number of two-photon interference (TPI) experiments have been conducted under various experimental conditions of an interferometric setup and with the employment of conceptually extended two-photon superposed states with highly correlated photons[16-18]. To date, HOM-type TPI experiments have also been performed by employing electrons[19], plasmons[20], bosonic atoms[21,22], phonons in trapped ions[23], spin waves[24], Rydberg excitations[25], and microwave-frequency photons[26], instead of optical photons. Other non-classical features of light interference have been experimentally observed in various types of interferometers such as the Mach-Zehnder, Michelson and Franson interferometers via the use of highly correlated photons[27-33]. In the meantime, many experiments have been carried out to investigate the classical analogues of the TPI[34-37].

Although the observations of non-classical or quantum effects in optical interference have sometimes been recognized as counterintuitive phenomena[5,6], in fact, the interference itself can be easier understood as the superposition of indistinguishable probability amplitudes in quantum mechanical terminology[38]. This "quantum intuition" allows us to distinguish multiphoton quantum interference effects from classical analogues[2,3,7], and therefore, it is possible to identify pure quantum effects that cannot be simulated with classical lights. In this study, we performed

TPI experiments with uncorrelated photons from a luminescent light source and compared our results with the expected ones for correlated photons. Further, we investigated the critical factors determining the TPI fringe pattern, such as the phase terms, visibilities, and spectral bandwidths related to individual single-photon and two-photon states. Our experimental results showed that the interference phenomenon itself exhibits the same features for both correlated and uncorrelated photons, at least in the single-photon counting regime.

## Results

**Interference of two uncorrelated photons.** When photons are fed into a Mach-Zehnder interferometer (MZI), as shown in Fig. 1, two types of number-correlated two-photon states are generated with equal probability in the two interferometer arms beyond the 50:50 beamsplitter (BS1). One is the state wherein the two input photons follow two different paths 1 and 2, expressed as $|2\rangle \xrightarrow{BS1} |1\rangle_1 |1\rangle_2$; therefore, the two photons exhibits no phase relations between each other in the two interferometer arms, and this two-photon state contributes to the HOM interference described as $|1\rangle_1 |1\rangle_2 \xrightarrow{BS2} 1/\sqrt{2}\left(|2\rangle_3 |0\rangle_4 + |0\rangle_3 |2\rangle_4\right)$. The other state corresponds to the case wherein the two photons are bunched together along the same path (1 or 2) beyond BS1, which can be described as $|2\rangle \xrightarrow{BS1} 1/\sqrt{2}\left(|2\rangle_1 |0\rangle_2 + e^{2i\phi}|0\rangle_1 |2\rangle_2\right)$; this path-entangled state is highly phase-sensitive in the interferometer. Here, note that the two input photons do not have to possess identical properties in terms of the internal degrees of freedom, such as polarization and wavelength[27-30,39]. Moreover, they can differ in terms of external degrees of freedom, such as the input spatial modes and the arrival time of the two incident photons at BS1[30,39,40].

To observe the HOM-type TPI with classical light such as coherent light[41-46] and low-coherence fluorescent light[47,48], a phase randomization has to be employed between the two interferometer arms in order to prevent the single-photon interference (SPI) effect as well as the interference of the path-entangled state[41,46]. In the classical-light HOM-type experiment, the maximum visibility bound limited to 50% is only partly considered to be due to the photon-number distribution of the light source itself. In fact, the limited visibility is the only cause of the coincidence-detection contribution arising from the path-entangled state, which is intrinsic for the two input photons only when not considering multi-photon inputs[46,49].

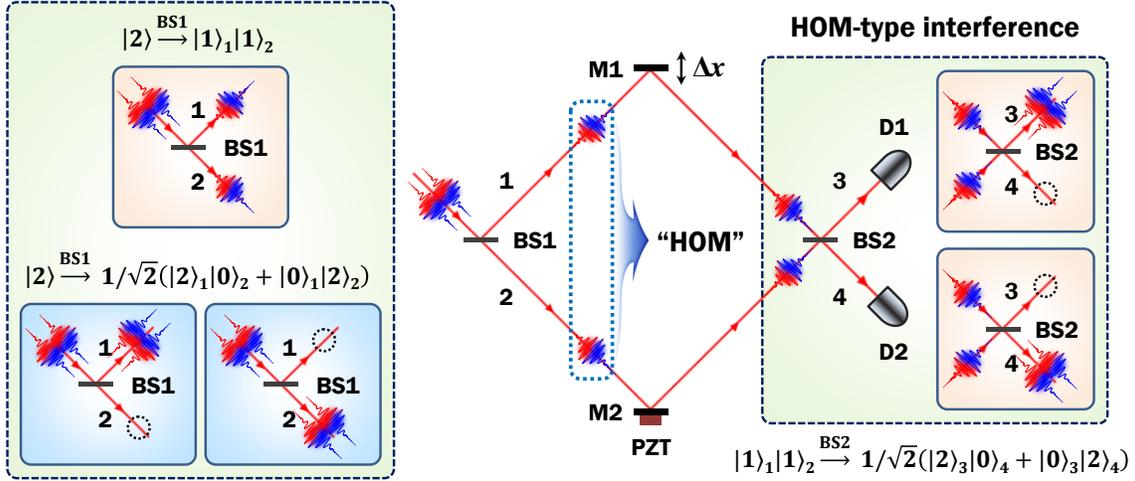

**Fig. 1** Mach-Zehnder interferometer employing two uncorrelated photons. When two photons are incident on BS1, two two-photon states are generated with equal probability within the two interferometer arms. One is the phase-insensitive separable state, which thus contributes to HOM-type interference, while the other is the highly phase-sensitive path-entangled state and thus shows a phase super-resolved interference fringe at the output port of BS2. BS, 50:50 beamsplitter; M, mirror; PZT, piezoelectric transducer; D, single-photon detector; $\Delta x$, introduced path-length difference between the two interferometer arms.

At the two MZI-output ports shown in Fig. 1, the single-photon counting rates recorded at the two single-photon detectors (SPDs) D1 and D2 after photon passage through the two interference filters IF1 and IF2 (not shown in Fig. 1), as a function of the path-length difference $\Delta x$, are given by

$$R_1 = R_{1,\infty}\left[1 - V_1 \cos(\phi_1) \mathbb{F}(\omega_1)\right], \quad R_2 = R_{2,\infty}\left[1 + V_2 \cos(\phi_2) \mathbb{F}(\omega_2)\right], \tag{1}$$

where $R_{i,\infty}$ represents the counting rate recorded at the two detectors located at the two output ports of BS2 for $\Delta x \gg l_{\text{coh.}}$ (where $l_{\text{coh.}}$ denotes the coherence length), $V_i$ denotes the SPI-fringe visibility, and $\phi_i = 2\pi \Delta x / \lambda_i$ (where, $\lambda_i$ denotes the center wavelength) represents the relative phase difference related to the path-length difference between the two interferometer arms. Further, $\mathbb{F}(\omega_i)$ denotes an interference-fringe-envelope function with central angular frequency $\omega_i$ and spectral bandwidth $\delta\omega_i$, which is related to the single-photon coherence length and thus actually determined by the spectral bandwidths of the filters used in the experiment[47]. The SPI effect expressed by Eq. (1) is the same as that in the cases of correlated and uncorrelated photons as well as in the interference of the classical electric fields of light.

However, in the case of the coincidence counting of uncorrelated photons with a very short coherence time (~picoseconds), the registered coincidences within the resolving time of electronics (~nanoseconds) are all accidental events. Therefore, the coincidence counting rate is given by

$$R_c = R_{c,\infty} \left\{ \begin{array}{l} 1 - V_1 \cos(\phi_1) \mathbb{F}(\omega_1) + V_2 \cos(\phi_2) \mathbb{F}(\omega_2) \\ -\dfrac{1}{2} V_1 V_2 \left[ \cos(\phi_1 + \phi_2) + \cos(\phi_1 - \phi_2) \right] \mathbb{F}(\omega_1) \mathbb{F}(\omega_2) \end{array} \right\}, \quad (2)$$

where $R_{c,\infty}$ denotes the coincidence counting rate recorded at two detectors for $\Delta x \gg l_{\text{coh.}}$. Actually, $R_c$ corresponds to the accidental coincidence given by $R_1 R_2 T_R$, where $T_R$ denotes the resolving time of the coincidence electronics. We note here that the TPI fringe represented by Eq. (2) exhibits a somewhat complex pattern and thus implies certain significant characteristics: (i) In general, there is a fundamental difference between valid and accidental coincidences in which correlated and uncorrelated photons, respectively, are employed. Therefore, the coincidence counting rate is specifically determined by single-photon counting rates $R_1$ and $R_2$ measured at the two detectors D1 and D2 in Fig. 1 and also by the resolving time of the coincidence electronics. (ii) The TPI fringe expressed by Eq. (2) includes three phase-related terms, which are the term corresponding to the interference of the individual single-photons and the sum- and difference-frequency oscillation terms. The $\cos(\phi_1 + \phi_2)$ term corresponds to the phase super-resolved fringe arising due to the path-entangled state, while the $\cos(\phi_1 - \phi_2)$ term corresponds to the spatial-beating fringe by the frequency-entangled state. (iii) Equation (2) also includes two SPI-fringe-envelope terms and a TPI-fringe-envelope term in which the entire shape is determined only by the single-photon spectral bandwidth. For two uncorrelated photons, the width of the TPI-fringe envelope including both the sum- and difference-frequency oscillations is given by the product of the individual single-photon bandwidths, because the path-entangled two-photon states are also generated probabilistically from the single-photon wavepacket. In contrast, for correlated photons, the difference-frequency term still affects the extent of the SPI-fringe-envelope, but the range of the TPI-fringe corresponding to sum-frequency oscillation is determined by the phase-matching characteristics for two-photon generations[28-31]. This interferometric feature may form the most significant difference between correlated and uncorrelated photons[29]. (iv) In contrast to naive intuition, it is interesting that the TPI-fringe pattern is strongly affected by the two SPI-fringes and also by their visibilities, as can be deduced from Eq. (2). Moreover, the TPI-fringe visibility corresponding to the sum- and difference-frequency oscillations is given as the product of two SPI-fringe visibilities. Therefore, the TPI effect corresponding to the super-resolution-fringe period disappears very rapidly relative to the reduction in the SPI-fringe visibility. Here, it is necessary to emphasize that the TPI-fringe expressed by Eq. (2) has the same pattern as that resulting from the use of two highly correlated photons in the experiment[29]. Furthermore, the TPI-fringe pattern in Eq. (2) can also be expressed by classical intensity-correlation measurement[41,50].

To simplify the analysis of the TPI-fringe full-width, we assume that the two filters have a Gaussian shape; consequently, the TPI-fringe envelope function can be expressed as $\mathbb{F}(\omega_1)\mathbb{F}(\omega_2) = \exp(-\Delta x^2/\sigma_{12}^2)$, where $\sigma_{12} = \sqrt{2\sigma_1^2\sigma_2^2/(\sigma_1^2+\sigma_2^2)}$ is related to the average spectral bandwidth ($\sigma_i = c/\delta\omega_i$, where $c$ denotes the speed of light in vacuum) of the two filters and determines the range of the TPI-fringe showing sum- and difference-frequency oscillations. In the case of $\Delta x \approx 0$, the TPI fringe corresponds to a super-resolved period corresponding to the sum-frequency oscillation, whereas in the case of $\Delta x > \sqrt{2}\sigma_{12}$, the oscillation period corresponds to the average value of the individual single-photon wavelengths. If the two interferometer arms are phase randomized by means of the piezoelectric transducer (PZT) in Fig. 1, then the TPI effect expressed by Eq. (2) corresponds to the spatial-beating fringe by the frequency-entangled state of two uncorrelated photons, wherein the fringe pattern

"contains" the $\cos(\phi_1 - \phi_2)$ term only. When the two filters have the same bandwidth, the spatial-beating-fringe width becomes narrower by $\sqrt{2}$ than that of the SPI ($\sigma_{12} = \sqrt{2}\sigma_{1,2}$).

If we now consider the most simplified case in which two identical photons are employed in the TPI experiment, Eq. (2) can be expressed as the form

$$R_c = R_{c,\infty}\left\{1-(V_1-V_2)\cos(\phi)\mathbb{F}(\omega)-\frac{1}{2}V_1V_2\left[1+\cos(2\phi)\right]\mathbb{F}(\omega)^2\right\}, \quad (3)$$

which reveals a super-resolution-fringe period across the entire range of the TPI-fringe irrespective of the non-vanishing SPI effect except for the fact that an asymmetric fringe oscillation is generated only for the case of $V_i \neq 0$. When the two SPI fringes show the same visibility, $V_1 = V_2$, the two SPI terms disappear and thus, Eq. (3) simply represents the sum of the HOM interference and the path-entangled-state interference fringes. Even in this case, it is interesting that the TPI fringe visibility is given by the product of the two SPI-fringe visibilities, regardless of whether the fringes are actually observed in individual detectors, particularly for HOM interference.

**Experimental setup.** Figure 2 shows the experimental setup utilized to perform TPI experiments employing uncorrelated photons from a weak and incoherent light source. In the study, we used single-mode-fiber (SMF)-coupled broadband light from a super-luminescent diode (Qphotonics, QSDM-810), which has a central wavelength of 810 nm and spectral bandwidth of 23.5 nm. In the setup, highly attenuated light to the single-photon level is horizontally polarized by PBS1 and then fed into a polarization-based Michelson interferometer. The two polarization components are equally divided by PBS2 after photon passage through a half-wave plate (HWP1) with its axis oriented at 22.5°. A relative path-length difference $\Delta x$ between the two interferometer arms is introduced by moving the mirror M1 mounted on a translation stage. Active relative-phase randomization between the two interferometer arms is performed by use of the PZT actuator, which is utilized essentially to observe the phase-insensitive HOM interference fringes. The PZT actuator is also used to precisely measure interference fringes. Two quarter-wave plates (QWPs) with their axes oriented at 45° are placed in the two interferometer arms to rotate the polarization direction. The output photons from PBS2 are passed through HWP2 with its axis oriented at 22.5° and thence to PBS3. Finally, the two output photons from PBS3 are coupled into two SMFs through the two interference filters (IF1 and IF2). In our experiment, one of the filters (IF2) was tilted at an angle of 20° with respect to the normal direction to shift the central wavelength of the transmitted spectrum of IF2. The central wavelengths of the two filters IF1 and IF2 are 810.63 nm and 798.44 nm, respectively. Two SPDs are connected to two fiber couplers (FCs) via SMFs, and subsequently, the output signals are fed into the single- and coincidence-counting electronics. In our experiment, the coincidence window was set to 10 ns, and thus, the average photon-number per resolving time was approximately $3\times10^{-3}$ at individual detectors D1 and D2 (not shown in Fig. 2). The operating principle of the polarization-based Michelson interferometer is the same as that of the MZI shown in Fig. 1[46].

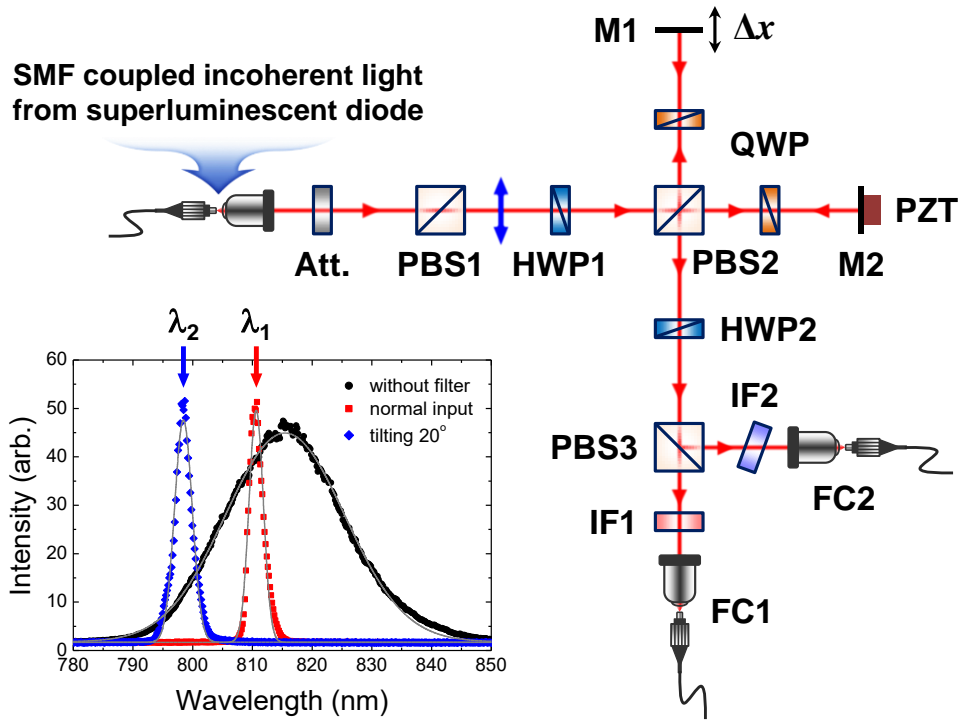

**Fig. 2** Experimental setup. A polarization-based Michelson interferometer is used to perform two-photon interference (TPI) experiments with uncorrelated photons. SMF, single-mode fiber; Att., attenuator; PBS, polarizing beam splitter; HWP, half-wave plate; QWP, quarter-wave plate; M, mirror; PZT, piezoelectric transducer; IF, interference filter; FC, single-mode-fiber coupler. Inset shows the two transmission spectra of the two IFs (see text for details).

In our experiment, we measured phase-sensitive SPI and TPI fringes without phase randomization of the interferometer in order to confirm the full-fringe shape and the effect of the phase-related terms in Eq. (2), which can reveal the characteristic features of the TPI fringe of uncorrelated photons. Here, we utilized two methods for detecting the two-photon coincidence events; one is the conventional coincidence counting with two SPDs present at the two interferometer output ports, and the other is the time-delayed coincidence measurement of successive electrical signals from each SPD, which is possible when the two input photons are temporally well-separated relative to the dead time of the SPD[46]. We also investigated the influence of the SPI-fringe visibility on the TPI-fringe visibility caused by two non-identical SPI-fringe visibilities. Furthermore, the phase-insensitive interference fringes were also measured upon applying the phase randomization.

**Experimental results.** Our experimental results are shown in Fig. 3. The SPI and TPI fringes are acquired with a 200-nm resolution of $\Delta x$. The measured single-photon counting rates are normalized (symbols) and compared with the theoretical predictions with the same parameters (very dense red lines). Here, the fringe-envelope function is considered as a Gaussian. Figures 3(a) and (3b) show the SPI fringes measured at individual detectors D1 and D2, respectively, which have visibilities and Gaussian-shaped fringe widths of $V_1 = 0.98$ and $V_2 = 0.90$, and $\sigma_1 = 0.17$ mm and $\sigma_2 = 0.08$ mm, respectively. Here, we note that asymmetric and non-ideal visibilities can arise due to the imperfect extinction ratio of PBS3 and also imperfect alignment of the interferometric setup. In our study, when IF2 was tilted at an angle of 20°, the transmitted spectrum was slightly changed from the Gaussian shape. As a result, unwanted side peaks appeared, as shown in Fig. 3(b); these peaks caused by rather complex spectral components can affect the TPI fringe shape. The insets in the figures represent the transmission spectra of each filter and the phase-resolved SPI fringes measured at $\Delta x \approx 0$.

To observe the TPI fringe with only one SPD, we utilized the time-delayed coincidence measurement of successive electrical signals from each detector. In this case, the two photons measured via the interferometer arrive at the same detector with a long time interval relative to the dead time of the SPD. Here, we remark that it is known that two temporally well-separated pairwise two-photon states show the same interferometric feature as in the case wherein the two photons are incident on the interferometer simultaneously[16,17,39,46]. Further, we have recently demonstrated that the HOM-type TPI effect can be observed with only one SPD by means of a time-delayed coincidence measurement when temporally well-separated pairwise weak coherent pulses are employed in the TPI experiment[46].

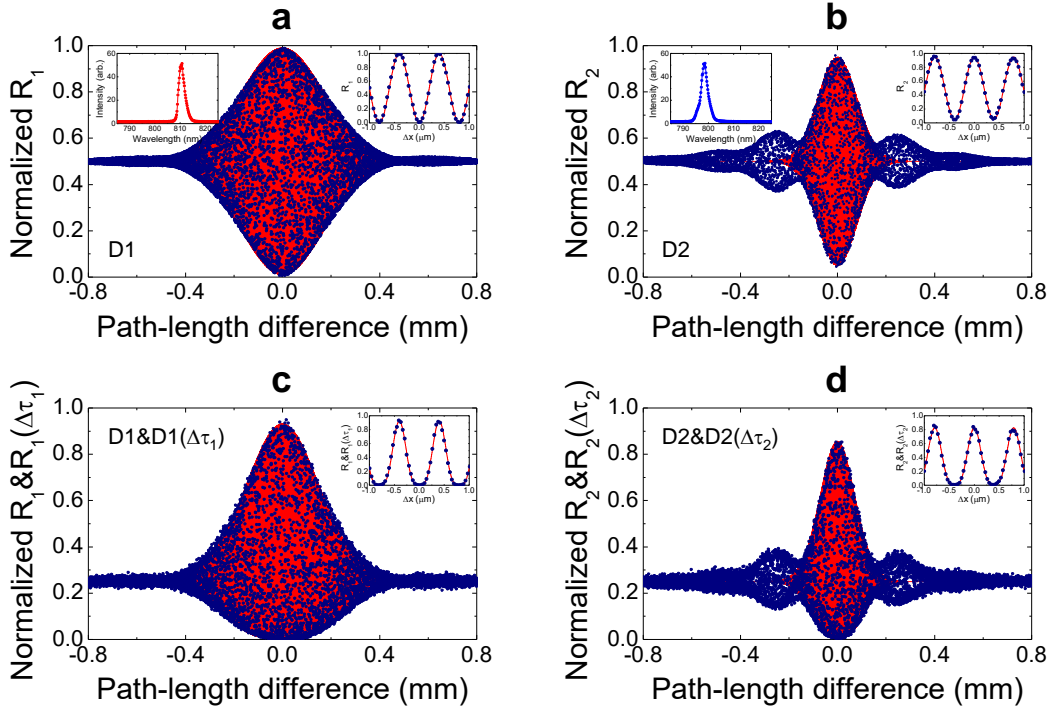

**Fig. 3** Experimental results for single-photon interference (SPI) and two-photon interference (TPI) of uncorrelated photons measured with one detector. (**a,b**) Normalized single counting rates measured at D1 and D2 as a function of the path-length difference showing visibilities of $V_1 = 0.98$ and $V_2 = 0.90$, respectively, for a coarse scan over the range of the full-fringe width. For $\Delta x \approx 0$, phase-resolved SPI fringes are obtained with visibilities $V_1 = 0.99 \pm 0.02$ and $V_2 = 0.90 \pm 0.01$ (see insets). (**c,d**) Normalized coincidence counting rates at D1&D1($\Delta \tau_1$) and D2&D2($\Delta \tau_2$) showing visibilities $V_1 = 0.93$ and $V_2 = 0.85$, respectively (see text for details). Phase-resolved TPI fringes are obtained with visibilities $V_1 = 0.94 \pm 0.02$ and $V_2 = 0.83 \pm 0.02$ (see insets). The red backgrounds represent the theoretical predictions.

The experimental result in this work confirms that this kind of measurement technique can also be applied to a continuous-mode weak incoherent light source. The TPI fringe measured at D1&D1($\Delta\tau_1$) or D2&D2($\Delta\tau_2$) can be expressed as

$$R_c = R_{c,\infty}\left\{1 \pm 2V_i \cos(\phi_i)\mathbb{F}(\omega_i) + \frac{1}{2}V_i^2\left[1+\cos(2\phi_i)\right]\mathbb{F}(\omega_i)^2\right\}. \tag{4}$$

Here, $R_c$ can be expressed as $R_i R_i(\Delta\tau_i)T_R$, as in Eq. (2), regardless of the time delay involved. Figures 3(c) and 3(d) show the TPI fringes measured with each detector followed by electrical delay lines with $\Delta\tau_1 = \Delta\tau_2 = 60$ ns. The normalized coincidences measured at D1&D1($\Delta\tau_1$) and D2&D2($\Delta\tau_2$) as a function of the path-length difference, $\Delta x$, show visibilities of $V_1 = 0.93$ and $V_2 = 0.85$, respectively. From the phase-resolved TPI fringes (see insets), the visibilities are found to be $V_1 = 0.94 \pm 0.02$ and $V_2 = 0.83 \pm 0.02$. Different from the SPI fringe, the effect of the reduced visibility is remarkably reflected in the constructive positions than the destructive positions of the interference fringes.

Figure 4 shows the experimental results for the TPI fringe measured with the two SPDs D1 and D2. Similar to Fig. 3, the normalized coincidence as a function of $\Delta x$ shows visibilities of $V_1 = 0.98$ and $V_2 = 0.90$ (red background). Here, the TPI fringe of uncorrelated photons is determined by envelope functions $\mathbb{F}(\omega_1)$, $\mathbb{F}(\omega_2)$, and $\mathbb{F}(\omega_1)\mathbb{F}(\omega_2)$ for each single-photon spectral bandwidth, as well as the two SPI-fringe visibilities $V_1$ and $V_2$ in Eq. (2). Figures 4(b) and 4(c) depict the phase-resolved fringes for $\Delta x \approx 0$ to demonstrate the characteristic features of nondegenerate photons showing the spatial-beating fringe involving the sum- and difference-frequency oscillation terms in Eq. (2) and to clarify the effect of the asymmetrical visibilities of the two SPI fringes on the TPI fringe. The two dark-red dotted lines in Fig. 4(b) indicate the constructive-interference positions involving sum-frequency oscillation, and the extended lines indicate the beat-fringe period corresponding to the difference-frequency oscillation term. The influence of the two non-identical SPI-fringe visibilities in Figs. 3(a) and 3(b) is reflected in the asymmetric fringe oscillation, as indicated by the light-blue lines in Figs. 4(b) and 4(c). The fringe visibilities in Fig. 4(c) are found to be $V_1 = 0.99 \pm 0.01$ and $V_2 = 0.89 \pm 0.01$, which agree well with the values indicated in Figs. 3(a) and 3(b). The effect of SPI on the TPI fringe is also shown in the case where highly correlated photons are involved in the interference experiments[29], although the SPI fringes are actually not observed in individual detectors[27,28,31].

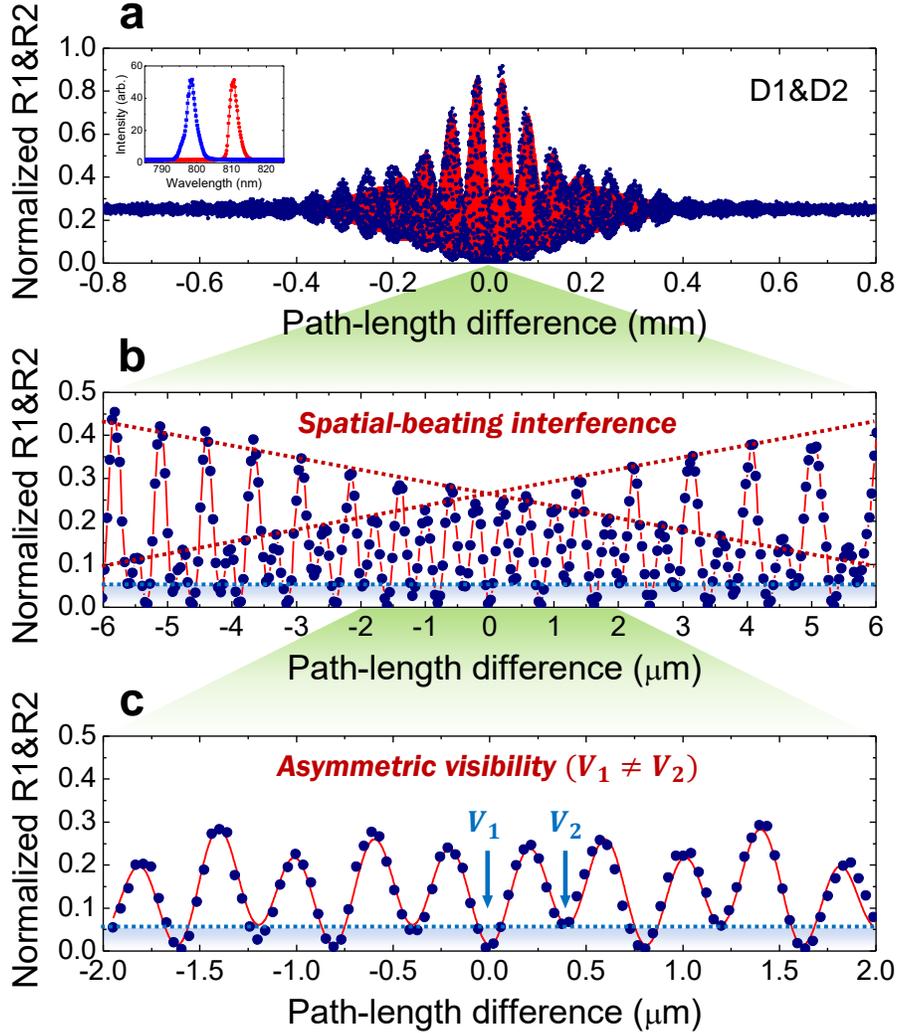

**Fig. 4** Experimental results for two-photon interference (TPI) of uncorrelated photons measured with two detectors. **a** Normalized coincidence counting rate measured at D1&D2 as a function of the path-length difference showing visibilities $V_1 = 0.99$ and $V_2 = 0.90$, for a coarse scan over the range of the full-fringe width. **b** and **c** correspond to a phase-resolved TPI fringe for $\Delta x \approx 0$, which is obtained with visibilities $V_1 = 0.99 \pm 0.01$ and $V_2 = 0.89 \pm 0.01$ as shown in **c**. The inset of **a** depicts the two transmission spectra of the two filters IF1 and IF2.

**Observing TPIs of uncorrelated photons with phase randomization.** Finally, we observed HOM-type TPI fringes upon employing relative-phase randomization between the two interferometer arms. In this case, the phase-related terms in Eq. (2) and Eq. (4) are canceled out, and consequently, the TPI fringes measured at D1&D2, D1&D1($\Delta\tau_1$), and D2&D2($\Delta\tau_2$) are expressed in the form

$$R_c = R_{c,\infty}\left\{1 - \frac{1}{2}V_1V_2\left[\cos(\phi_1 - \phi_2)\right]\mathbb{F}(\omega_1)\mathbb{F}(\omega_2)\right\}, \text{ for D1\&D2,}$$

$$R_c = R_{c,\infty}\left[1 + \frac{1}{2}V_i^2\mathbb{F}(\omega_i)^2\right], \text{ for D1\&D1}(\Delta\tau_1) \text{ and D2\&D2}(\Delta\tau_2). \quad (5)$$

Here, note that the visibilities corresponding to Eq. (5) are the same as those for Eq. (2) and Eq. (4), and thus governed by the SPI-fringe visibilities of Eq. (1), even though the SPI fringes are actually not observed in the individual detectors due to the random phase. The spatial-beating fringe measured at D1&D2 reveals the same feature as the TPI fringe of the frequency-entangled state formed with highly correlated photons[39,51-53], except for the limited visibility due to the path-entangled-state contributions. For coincidence measurement with one detector involving an electrical time delay, the HOM-type-peak fringe measured at D1&D1($\Delta\tau_1$) or D2&D2($\Delta\tau_2$) is similar to the case of coincidence measurement when two detectors are placed at one of the two output ports of the interferometer. The phase-insensitive peak fringes measured at D1&D1($\Delta\tau_1$) and D2&D2($\Delta\tau_2$) are shown in Fig. 5(a) and 5(b), respectively, which indicate visibilities of $V_1 = 0.90 \pm 0.01$ and $V_2 = 0.83 \pm 0.01$ and Gaussian widths of $\sigma_1 = 171.69 \pm 1.60$ μm and $\sigma_2 = 84.53 \pm 1.13$ μm. Figure 5(c) depicts the spatial-beating fringe measured at D1&D2 with visibilities of $V_1 = 0.97 \pm 0.01$ and $V_2 = 0.83 \pm 0.01$. From the measured results, we find the beat-fringe period to be $52.13 \pm 0.10$ μm, corresponding to the angular-frequency difference $\Delta\omega = 3.62 \pm 0.01 \times 10^{13}$ rad/s, which is well-matched with the value estimated from the center-wavelength difference of the two filters used in the experiment.

Here, we note that the measured HOM-type-peak fringe in one detector and the spatial-beating fringe in the two detectors have to be considered as genuine two-photon interference effects. In the TPI experiments with uncorrelated photons in the phase-randomized interferometer, the coincidence events registered within the resolving time do not exceed the accidental coincidences determined by the single-photon counting events in individual detectors. Therefore, the normalized coincidences exceeding 0.5, as shown in Fig. 5, must be obviously originate from the HOM-type and spatial-beating interferences of the two two-photon states at the single-photon level, $1/\sqrt{2}\left(|1\rangle_1|1(\Delta t)\rangle_2 + |1(\Delta t)\rangle_1|1\rangle_2\right)$ and $1/\sqrt{2}\left(|\omega_1\rangle_1|\omega_2\rangle_2 + |\omega_2\rangle_1|\omega_1\rangle_2\right)$, respectively. If we consider these two-photon states only, the interference phenomenon itself demonstrates the same feature as in the case when highly correlated photons are employed in the same interference experiment. This is because the TPI effect itself originates from the related two-photon states but not from the types of light sources used in the interference experiments.

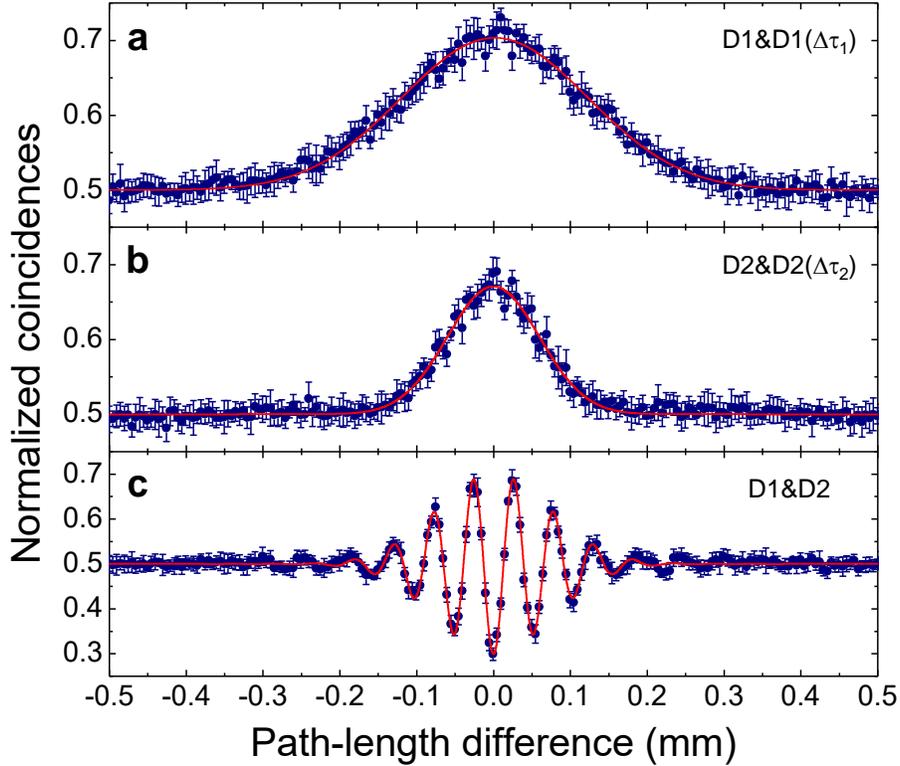

**Fig. 5** Experimental results for Hong-Ou-Mandel (HOM)-type two-photon interferences (TPIs) of phase-randomized uncorrelated photons. **a**, **b**, and **c** correspond to TPI fringes measured at D1&D1($\Delta\tau_1$), D2&D2($\Delta\tau_2$), and D1&D2, respectively.

**Discussion**

We have experimentally demonstrated the TPI of nondegenerate uncorrelated photons from a broadband super-luminescent light source. In the TPI experiments, the essential difference between correlated and uncorrelated photons is the valid coincidence events within the resolving time of the coincidence detection device and accidental coincidences, respectively. For uncorrelated photons employed in the TPI experiment, the observed TPI-fringe shape and visibility were fully determined by the two SPI-fringes and also by the visibilities, which have the same interferometric feature as in the case when highly correlated photons are used in the experiment within the range of the single-photon coherence length. The essential difference of the interferometric characteristics between correlated and uncorrelated photons is the interference of the path-entangled two-photon state revealing sum-frequency oscillations far beyond the coherence length of the source. Here, we have focused on two kinds of two-photon quantum states generated within the interferometer, which result in the phase-insensitive HOM-type interference and the interference of the highly phase-sensitive path-entangled state. When phase randomization was introduced in the interferometer arms, the limited HOM-type TPI visibility originated from the contribution of the constant coincidence events due to the path-entangled state. Furthermore, the normalized coincidence exceeding 0.5 obviously originated from the HOM-type and spatial-beating interferences of the corresponding two-photon states. With the exception of essential differences such as the number of coincidence events and the two-photon spectral bandwidth, the interference phenomenon itself reveals the same features for both the correlated and uncorrelated photons in the single-photon counting regime. Although the observed interference effects in our work are essentially arise from the two-photon states in the interferometer, the observed interference effects can also be explained by classical intensity-correlation measurements.

**References**

1. Kipnis, N. *History of the principle of interference of light* (Birkhäuser, 1991).
2. Greenberger, D. M., Horne, M. A. & Zeilinger, A. Multiparticle interferometry and the superposition principle. *Phys. Today* **46**, 22-29 (1993).
3. Pan, J. W. *et al.* Multiphoton entanglement and interferometry. *Rev. Mod. Phys.* **84**, 777-838 (2012).
4. Young, T. Experiments and calculations relative to physical optics. *Phil. Trans. R. Soc. Lond.* **94**, 1-16 (1804).
5. Mandel, L. Quantum effects in one-photon and two-photon interference. *Rev. Mod. Phys.* **71**, S274-282 (1999).
6. Zeilinger, A. Experiment and the foundations of quantum physics. *Rev. Mod. Phys.* **71**, S288-297 (1999).
7. Jaeger, G. & Sergienko, A. V. Multi-photon quantum interferometry, in *Progress in Optics 42* (ed Wolf, E.) 277-324 (Elsevier, 2001).
8. Steinberg, A. M, Kwiat, P. G. & Chiao, R. Y. Quantum optical tests of the foundations of physics in *Springer Handbook of Atomic, Molecular, and Optical Physics* (ed. Drake, G. W. F.) 1185-1213 (Springer, 2006).
9. Kok, P. *et al*. Linear optical quantum computing with photonic qubits. *Rev. Mod. Phys.* **79**, 135-174 (2007).
10. O'Brien, J. L., Furusawa, A. & Vučković, J. Photonic quantum technologies, *Nat. Photonics* **3**, 687-695 (2009).
11. Gisin, N. & Thew, R. Quantum communication. *Nat. Photon*. **1,** 165-171 (2007).
12. Giovannetti, V., Lloyd, S. & Maccone, L. Quantum-enhanced measurements: beating the standard quantum limit. *Science* **306**, 1330-1336 (2004).
13. Yuan, Z. S. *et al*. Entangled photons and quantum communication. *Phys. Rep*. **497**, 1-40 (2010).
14. Degen, C. L., Reinhard, F. & Cappellaro, P. Quantum sensing. *Rev. Mod. Phys*. **89**, 035002 (2017).
15. Hong, C. K., Ou, Z. Y. & Mandel, L. Measurement of subpicosecond time intervals between two photons by interference. *Phys. Rev. Lett.* **59**, 2044-2046 (1987).

16. Pittman, T. B. *et al*. Can two-photon interference be considered the interference of two photons?. *Phys. Rev. Lett.* **77**, 1917-1920 (1996).

17. Kim, Y. -H. & Grice, W. P. Quantum interference with distinguishable photons through indistinguishable pathways. *J. Opt. Soc. Am. B* **22**, 493-498 (2005).

18. Kim, H., Lee, S. M. & Moon, H. S. Generalized quantum interference of correlated photon pairs. *Sci. Rep.* **5**, 9931 (2015).

19. Bocquillon, E. *et al*. Coherence and indistinguishability of single electrons emitted by independent sources. *Science* **339**, 1054-1057 (2013).

20. Fakonas, J. S., Lee, H., Kelaita, Y. A. & Atwater, H. A. Two-plasmon quantum interference. *Nat. Photonics* **8**, 317-320 (2014).

21. Lopes, R. *et al.* Atomic Hong–Ou–Mandel experiment. *Nature* **520**, 66-68 (2015).

22. Kaufman, A. M. et al. Two-particle quantum interference in tunnel-coupled optical tweezers. *Science* **345**, 306-309 (2014).

23. Toyoda, K., Hiji, R., Noguchi, A. & Urabe, S. Hong-Ou-Mandel interference of two phonons in trapped ions. *Nature* **527**, 74-77 (2015).

24. Parniak, M. et al. Quantum optics of spin waves through ac Stark modulation. *Phys. Rev. Lett*. **122**, 063604 (2019).

25. Li, J. et al. Hong-Ou-Mandel interference between two deterministic collective excitations in an atomic ensemble. *Phys. Rev. Lett.* **117**, 180501 (2016).

26. Lang, C. et al. Correlations, indistinguishability and entanglement in Hong–Ou–Mandel experiments at microwave frequencies. *Nat. physics* **9**, 345-348 (2013).

27. Ou, Z. Y., Zou, X. Y., Wang, L. J. & Mandel, L. Experiment on nonclassical fourth-order interference. *Phys. Rev. A* **42**, 2957-2965 (1990).

28. Rarity, J. G. *et al*. Two-photon interference in a Mach-Zehnder interferometer. *Phys. Rev. Lett*. **65**, 1348-1351 (1990).

29. Larchuk, T. S. *et al*. Interfering entangled photons of different colors. *Phys. Rev. Lett*. **70**, 1603-1606 (1993).

30. Shih, Y. H., Sergienko, A. V., Rubirin, M. H., Kiess, T. E. & Alley, C. O. Two-photon interference in a standard Mach-Zehnder interferometer. *Phys. Rev. A* **49**, 4243-4246 (1994).

31. Edamatsu, K., Shimizu, R. & Itoh, T. Measurement of the photonic de Broglie wavelength of entangled photon pairs generated by spontaneous parametric down-conversion. *Phys. Rev. Lett*. **89**, 213601 (2002).

32. Brendel, J., Mohler, E. & Martienssen, W. Time-resolved dual-beam two-photon interferences with high visibility. *Phys. Rev. Lett*. **66**, 1142-1145 (1991).

33. Ou, Z. Y., Zou, X. Y., Wang, L. J. & Mandel, L. Observation of nonlocal interference in separated photon channels. *Phys. Rev. Lett*. **65**, 321-324 (1990).

34. Kaltenbaek, R., Lavoie, J. & Resch, K. J. Classical analogues of two-photon quantum interference. *Phys. Rev. Lett.* **102**, 243601 (2009).

35. Kaltenbaek, R., Lavoie, J., Biggerstaff, D. N. & Resch, K. J. Quantum-inspired interferometry with chirped laser pulses. *Nat. Physics* **4**, 864-868 (2008).

36. Ogawa, K., Tamate, S., Kobayashi, H., Nakanishi, T. & Kitano, M. Time-reversed two-photon interferometry for phase superresolution. *Phys. Rev. A* **88**, 063813 (2013).

37. Ogawa, K., Tamate, S., Nakanishi, T., Kobayashi, H. & Kitano, M. Classical realization of dispersion cancellation by time-reversal method. *Phys. Rev. A* **91**, 013846 (2015).

38. Feynman, R. P., Leighton, R. B. & Sands, M. Quantum Mechanics *in Lectures on Physics,* (Addison-Wesley, 1965).

39. Kim, H., Lee, S. M. & Moon, H. S. Two-photon interference of temporally separated photons. *Sci. Rep*. **6**, 34805 (2016).

40. Jin, X. -M. *et al*. Sequential path entanglement for quantum metrology. *Sci. Rep.* **3**, 1779 (2013).

41. Kim, Y. S., Slattery, O., Kuo, P. S. & Tang, X. Conditions for two-photon interference with coherent pulses, *Phys. Rev. A* **87**, 063843 (2013).

42. Kim, Y. S., Slattery, O., Kuo, P. S. & Tang, X. Two-photon interference with continuous-wave multi-mode coherent light, *Opt. Express* **22**, 3611-3620 (2014).

43. Wang, C. *et al*. Realistic device imperfections affect the performance of Hong-Ou-Mandel interference with weak coherent states, *J. Lightwave Technol.* **35**, 4996-5002 (2017).

44. Chen, H. *et al*. Hong–Ou–Mandel interference with two independent weak coherent states, *Chin. Phys. B* **25**, 020305 (2016).

45. Moschandreou, E. *et al*. Experimental study of Hong–Ou–Mandel interference using independent phase randomized weak coherent states, *J. Lightwave Technol.* **36**, 3752-3759 (2018).

46. Kim, H., Lee, S. M., Kwon, O. & Moon, H. S. Observation of two-photon interference effect with a single non-photon-number resolving detector. *Opt. Lett.* **42**, 2443-2446 (2017).

47. Ou, Z. Y., Gage, E. C., Magill, B. E. & Mandel, L. Fourth-order interference technique for determining the coherence time of a light beam. *J. Opt. Soc. Am. B* **6**, 100-103 (1989)

48. Baba, M., Li, Y. & Matsuoka, M. Intensity Interference of ultrashort pulsed fluorescence. *Phys. Rev. Lett.* **76**, 4697-4700 (1996).

49. Aragoneses, A. *et al*. Bounding the outcome of a two-photon interference measurement using weak coherent states. *Opt. Lett*. **43**, 3806-3809 (2018).

50. Rarity, J. G., Tapster, P. R. & Loudon, R. Non-classical interference between independent sources, *J. Opt. B: Quantum Semiclass. Opt*. **7**, S171-S175 (2005).

51. Ou, Z. Y. & Mandel, L. Observation of spatial quantum beating with separated photodetectors. *Phys. Rev. Lett.* **61**, 54-57 (1988).

52. Kim, H., Ko, J. & Kim, T. Two-particle interference experiment with frequency-entangled photon pairs. *J. Opt. Soc. Am. B* **20**, 760-763 (2003).

53. Park, J., Kim, H. & Moon, H. S. Two-photon interferences of nondegenerate photon pairs from Doppler-broadened atomic ensemble. *Opt. Express* **25**, 32064-32073 (2017).

**Acknowledgements**

This work was supported by the Basic Science Research Program through the National Research Foundation of Korea funded by the Ministry of Education, Science and Technology (No. 2018R1A2A1A19019181 and No. 2016R1D1A1B03936222).

**Author Contributions**

H.K., O.K. and H.S.M. conceived the project. H.K. designed the experimental setup and performed the experiment. H.K., O.K. and H.S.M. discussed the results and contributed to the writing of the manuscript.